\documentclass[twocolumn,showpacs,preprintnumbers,amsmath,amssymb]{revtex4}

\usepackage{graphicx}
\usepackage{dcolumn}
\usepackage{bm}

\begin{document}

\preprint{APS/123-QED}

\title{Device-spectroscopy of magnetic field effects in a polyfluorene organic light-emitting diode}

\author{T. D. Nguyen, J. Rybicki, Y. Sheng, M. Wohlgenannt} \email{markus-wohlgenannt@uiowa.edu}

\address{Department of Physics and Astronomy and Optical Science and Technology Center, University of Iowa, Iowa City, IA 52242-1479, USA}

\date{\today}

\begin{abstract}
We perform charge-induced absorption and electroluminescence spectroscopy in a polyfluorene organic magnetoresistive device. Our experiments allow us to measure the singlet exciton, triplet exciton and polaron densities in a live device under an applied magnetic field, and to distinguish between three different models that were proposed to explain organic magnetoresistance. These models are based on different spin-dependent interactions, namely exciton formation, triplet exciton-polaron quenching and bipolaron formation. We show that the singlet exciton, triplet exciton and polaron densities and conductivity all increase with increasing magnetic field. Our data are inconsistent with the exciton formation and triplet-exciton polaron quenching models.
\end{abstract}


\maketitle

\section{\label{sec:Introduction}Introduction}

Recent years have seen a surge in interest in spin-transport in organic semiconductor devices, including the study of the organic magnetoresistive effect (OMAR). OMAR is a recently discovered large, low-field, room-temperature magnetoresistive effect in \emph{non-magnetic} organic light-emitting diodes (OLEDs) \cite{Francis:2004,Mermer:2005,Prigodin:2006}. The effect can be as large as $10\%$ relative change in resistance for a magnetic field $B=10$ mT. To the best of our knowledge, the mechanism causing OMAR is not yet established with certainty. Independently, the effect of small magnetic fields on spin-dynamics in electron-hole pairs have also been studied for several decades in organics (for a review see Ref.~ \cite{Steiner:1989}). The concepts that arose from these works in radical pairs have recently been applied to OLEDs: magnetic field effects (MFE) on photocurrent \cite{Frankevich:1992,Hu:2007}, electroluminescence (EL) \cite{Kalinowski:August2003,Hu:2007,Odaka:2006,Iwasaki:2006} and exciton dissociation at electrodes \cite{Kalinowski:August2003} were found experimentally and modeled based on a postulated spin-dependence between the singlet and triplet radical pair channels of recombination. In addition, the interaction between triplet excitons and radicals was studied in anthracene crystals by Ern and Merrifield \cite{Ern:1968}. The question naturally arises how relevant the concepts of radical pair and triplet-radical spin-dynamics are to spin-transport in organics. We will study this question in the context of OMAR.

Three kinds of models have been put forward to explain OMAR. (i) Electron-hole pair mechanism (EHP) models based on concepts borrowed from the before mentioned MFE in radical pairs \cite{Frankevich:1992,Kalinowski:August2003,Hu:2007,Prigodin:2006,Odaka:2006,Iwasaki:2006}. In this model the spin-dependent reaction $P^+ + P^-\rightarrow {\rm exciton}$ between oppositely charged polarons to form an exciton ("recombination") is of central importance. (ii) The triplet-exciton polaron quenching model \cite{Desai:2007} (TPQ) that is based on the spin-dependent reaction $TE+P\rightarrow P+GS^*$ between a triplet exciton (TE) and a polaron to give an excited singlet ground state ($GS^*$). (iii) The bipolaron mechanism (BP) \cite{Wohlgenannt:2006} which treats the spin-dependent formation of doubly occupied sites (bipolarons) $P^+ + P^+ \rightarrow BP^{2+}$ (and an analogous reaction for negative carriers) during the hopping transport through the organic film. We note that the BP model does not assume the formation of stable bipolarons, but is merely based on the occurrence of doubly occupied hopping sites whose energy may be higher than that of two singly occupied sites. All three models are based on spin-dynamics induced by the hyperfine interaction \cite{Schulten:1978}. We note that a phenomenological model similar to BP was used by Movaghar {\emph et al.} \cite{Movaghar:1978} to explain the relatively small positive and negative magnetoresistance \cite{Clark:1974,Mell:1970} in amorphous inorganic semiconductors. Before further progress in the understanding of magnetotransport in organic semiconductors can be made, experiments must be completed to distinguish between this three directions. First, we note that the EHP and TPQ models require exciton formation (and therefore the presence of both majority and minority carriers), whereas BP can exist also in unipolar devices. Sheng \emph{et al.} \cite{Sheng:2006} found a significant dependence of the magnitude of OMAR on the minority carrier density, but this dependence is much weaker than the linear dependence that would be expected from EHP. Desai et al. \cite{Desai:2007} found that the onset of OMAR coincides with that of minority carrier injection, but Nguyen et al. \cite{Nguyen:2007} showed that a small OMAR also exists in doped polythiophene-derivatives, clearly a unipolar device. We therefore conclude that the experiments so far did not provide a definite answer. Here we therefore report on a different test that aims directly at observing the spin-dependent reactions.

In the present work, we put these models to a stringent test by measuring the dependence of the densities of the singlet, triplet excitons and polarons on the applied magnetic field, $B$. The three models make qualitatively different predictions for the MFE on these densities.

\section{Experimental}

 To measure the triplet and polaron densities we use the charge-induced absorption (CIA) spectroscopy technique under an applied B. In this experiment the changes in the device transmission spectrum resulting from induced absorption of the injected carriers and their recombination by-products is detected. The singlet excitons are detected using EL spectroscopy. For these measurements we used a standard OLED architecture as described in more detail previously \cite{Mermer:2005PRB}: A Poly (3,4-ethylenedioxythiophene)-poly (styrenesulfonate)(PEDOT) layer was spin-coated onto an indium-tin-oxide (ITO) covered glass slide as the hole-injecting electrode. The active polymer film (100nm), poly(9,9-dioctylfluorenyl-2,7-diyl)(PFO, see Fig.~\ref{fig:CIA} inset), purchased from American Dye Source (ADS), was spin coated onto the substrate from a chloroform solution with a concentration of 10 mg/ml. Finally, the Ca cathode (with an Al capping layer) was deposited to complete the sandwich structure. The sensitive manufacturing steps were carried out in a nitrogen glove box. A device area of $\approx 12$ mm$^2$ was chosen to match the size of the tungsten lamp filament used in CIA spectroscopy. 
 For the CIA measurement an applied AC voltage, modulated at about 1000 Hz between 0 V and a certain voltage level, was used resulting in a current of $\approx$ 1mA. A monochromatized tungsten-halogen lamp (250 W) was used as the probe beam
 . The 
 CIA spectra were obtained by plotting the negative fractional change in transmission, $-\Delta T/T$ versus the probe photon energy. $-\Delta T/T$ is proportional to the induced change in absorption coefficient. The MFE on CIA, $\delta T/\Delta T \equiv (\Delta T(B)-\Delta T(0))/\Delta T(0)$ and EL, $\delta(EL)/EL \equiv (EL(B)-EL(0))/EL(0)$ were measured under identical conditions together with the magnetoconductance.

\section{\label{sec:ExperimentalResultsI} Experimental Results:}

\begin{figure}
 \includegraphics[width=\columnwidth]{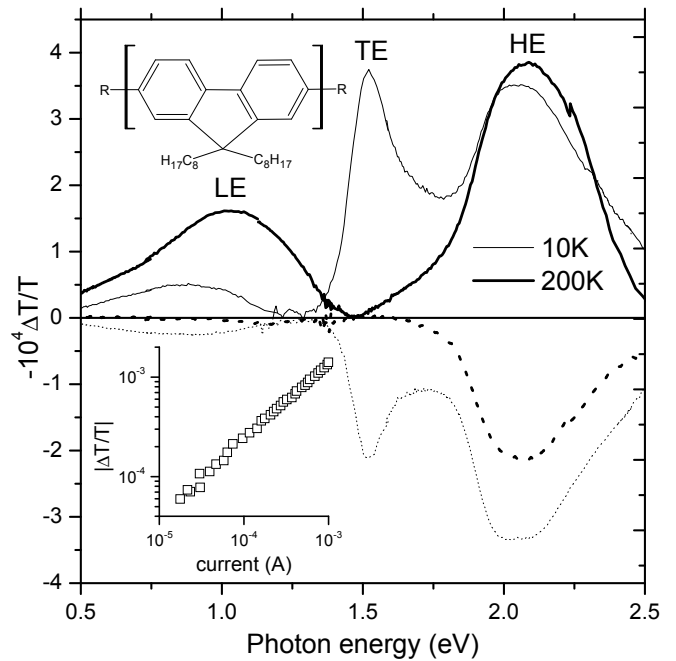}\\
 \caption{The in-phase (solid) and quadrature (dotted) CIA spectra of a PFO device (100nm thick) at 10 K (thin lines) and 200 K (bold lines) modulated at 1000 Hz. The triplet exciton (TE) and high and low energy (HE and LE, respectively) peaks are assigned. The insets show the dependence of the magnitude of the TE band on device current and the chemical structure of PFO.}
 \label{fig:CIA}
\end{figure}

\begin{figure}
 \includegraphics[width=\columnwidth]{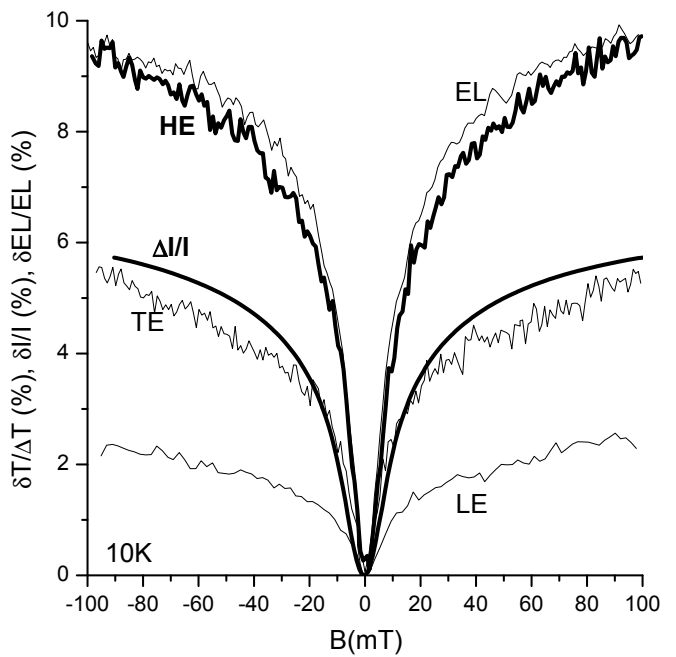}\\
 \caption{The magnetic field effect, $\delta T/\Delta T$ on the various CIA bands, current ($\delta I/I$) and EL ($\delta(EL)/EL$) at 10K.}
 \label{fig:MFE}
\end{figure}

Fig.~\ref{fig:CIA} shows the in-phase (solid) and quadrature (dotted) CIA spectra of a PFO device (100nm thick) at 10 K and 200K. Three absorption bands were observed, a narrow band at 1.52eV (TE) and two broader bands, a high-energy (HE) band at 2.1eV and a low energy (LE) band that shifts its peak position with temperature. The TE band is well-known to result from triplet exciton absorption \cite{Heyer:2005}. Dhoot {\emph et al.} \cite{Dhoot:2002} previously studied CIA in PFO devices similar to ours and reported a very similar spectrum. They "tentatively" assigned the $HE$ band to polaron absorption. The assignment of the $HE$ band to polarons has, to the best of our knowledge, not yet been established with certainty, but, since it is the dominant band in CIA where the primary effect is the injection of carriers, this assignment seems plausible. Furthermore, Dhoot {\emph et al.} \cite{Dhoot:2002} assigned LE also to polarons, since the HE and LE bands "were found to have the same dependence on voltage, driving frequency and temperature, indicating they belong to a single species." However, we obtained different results: Fig.~\ref{fig:CIA} shows that these two bands have a different ratio between in-phase and quadrature signals (and therefore a different lifetime) and a different temperature dependence. Furthermore, at relatively low currents we observe spectra that contain the HE band, but no LE band. These two bands therefore belong to two different species. The assignment of LE is therefore, to the best of our knowledge, not known at present.

Next we turn to the main experiment in this paper, the study of the MFE on the various bands in the CIA spectrum together with that of the singlet EL and the current. Fig.~\ref{fig:MFE} shows $\delta T/\Delta T$ measured at three fixed probe photon energies, namely 1.0 eV, 1.5 eV and 2.1eV. The magnetoconductivity, $\delta I/I$ and $\delta(EL)/EL$ are also shown. All five quantities increase with B (i.e. the fractional MFE is positive). Our results therefore show that the singlet exciton density (measured using EL), the triplet exciton density as well as the polaron density all simultaneously increase with B together with the device current. Since all these traces share the same dependence on $B$, they are clearly caused by a single common MFE. In agreement with our earlier results \cite{Mermer:2005}, $\delta(EL)/EL$ is somewhat larger than $\delta I/I$. $\delta T/\Delta T$ of TE is similar in magnitude as $\delta I/I$, whereas $\delta T/\Delta T$ of $HE$ is similar in magnitude as $\delta (EL)/EL$. $LE$ is significantly less magnetic field dependent than the other bands.

\section{Discussion} \label{sec:discussionpartI}

Next we separately discuss the experimental findings within the frame work of the three models. The interpretation of the experimental data shown in Fig.~\ref{fig:MFE} in an EHP-type model would occur along the following lines \cite{Sheng:2006}: In an OLED device, electrons and holes are injected from the cathode and anode, respectively, into the organic layer. During the recombination of free carriers into singlet and triplet excitons they form intermediate electron-hole pairs, either singlet pairs (SP) or triplet pairs (TP). SP finally form singlet excitons, which emit fluorescent light, and TP form triplet excitons, which are usually non-radiative. However, because of the negligible exchange interaction the spin states of SP and TP are mixed by the hyperfine interaction. In particular, in the absence of B, the singlet mixes with the entire triplet manifold. However, an applied magnetic field lifts the triplet degeneracy, and for a field strong compared to the hyperfine strength, the mixing remains only between the singlet and the triplet state with magnetic quantum number zero. It has been shown by several authors \cite{Frankevich:1992,Kalinowski:August2003,Prigodin:2006,Sheng:2006} that the application of $B$ therefore leads to enhanced singlet exciton formation together with reduced triplet exciton formation if the triplet exciton formation rate is larger than that of singlet excitons, and vice versa. In any case, one type of exciton is formed at the expense of the other, in clear disagreement with our experimental data (Fig.~\ref{fig:MFE}). Reufer. \emph{et al.} \cite{Reufer:2005} recently showed that in a (weakly) phosphorescent OLED the magnetic field dependence on singlet and triplet emission, i.e. electroflourescence and -phosphorescence, is equivalent and that both have a positive MFE, in agreement with our results. These authors also drew the conclusion that this observation disproves the EHP model as a candidate for explaining OMAR. However, we believe that their experiment is not 100\% conclusive, since in phosphorescent devices it is not possible to distinguish between phosphorescence that occurs as a result of intersystem crossing from a singlet exciton, or from a triplet exciton directly formed from polaron pairs.

In the TPQ model the mobility of polarons is affected by spin-dependent scattering processes obeying the reaction equation $TE+P\rightarrow P+GS^*$. Whereas the mobility of polarons is affected by this reaction, the number of polarons is conserved as is evident from the fact that both sides of the equation contain the same number of polarons. This is of course dictated by charge conservation: the number of polarons can only change as a result of either recombination with an oppositely charged polaron, or because of bipolaron formation between two equally charged polarons. Therefore our result that the polaron population changes with magnetic field is inconsistent with this model. There are other weaknesses of the TPQ model: It is based on a bimolecular reaction whose strength is proportional to the triplet exciton density times the polaron flow (current). As is shown in Fig.~\ref{fig:CIA}, inset, the triplet exciton density in the devices increases almost linearly (the fitted exponent is 0.77) with increasing current. The TPQ mechanism should therefore increase with increasing current, in contradiction with the experimental result that OMAR traces in many materials, e.g. in PFO (see Ref.~\cite{Mermer:2005PRB}, Fig. 1), actually decrease with increasing current. Furthermore, Fig.~\ref{fig:CIA} shows that the triplet density is measurably large only at low temperature (see also Ref.\cite{Dhoot:2002}, Fig. 2), in contradiction with the relative temperature insensitivity of OMAR \cite{Mermer:2005PRB}. This reduction of TE population with temperature is known to result from a decrease in intrinsic TE lifetime with increasing temperature, from roughly 10ms at 10K \cite{Coelle:2004} to $\approx 25\mu s$ at 300K \cite{Baldo:2000} in a representative material. We believe that these observations are all but impossible to reconcile with the TPQ model.

In a BP model, the MFE does not act on excitons, but affects the carrier mobility directly, leading, in the present case, to an increase in current. Obviously, this increase in current results in a corresponding increase in the by-products of carrier recombination \cite{Sheng:2006}, such as singlet and triplet excitons. We note that the magnitude of the MFE on singlet and triplet excitons need not necessarily be equal to that of the magnetoconductance \cite{Mermer:2005}, because carrier recombination is bimolecular. Furthermore, if the magnetoconductance of minority and majority carriers can be different, then the MFE on excitons and current need not show a simple relation to each other at all. However, the experimental observation that the MFE on the singlet excitons is about twice as large as that for the triplets appears at first sight to be at odds with the BP model, since in this model the excitons are assumed to form independently of their spin-state. However, it is straightforward to show (using a rate equation treatment) that this 1:2 ratio is expected for the scenario that singlet excitons recombine monomolecularly, whereas the triplet excitons recombine bimolecularly (the factor of two comes from the exponent of the bimolecular recombination term). That this is indeed the case is demonstrated by the observation that the continuous wave photoluminescence in organic films depends linearly on the laser power, whereas the triplet PA is proportional to the square-root of the laser power \cite{Wohlgenannt:2002}. Our findings are therefore in agreement only with the BP model. We note that our results obviously do not prove the validity of the BP model. For this it is probably necessary to show that the bipolaron density, if it can be detected, has a negative MFE.

\section{Conclusion}

We reported the charge-induced absorption spectrum of a polyfluorene OLED, and showed that it contained the signature of triplet excitons and polarons together with a low-energy band whose assignment is presently unknown. We measured the dependence of the densities of the singlet and triplet excitons, as well as polarons, on the applied magnetic field. Our results show that the singlet exciton, triplet exciton and polaron densities simultaneously increase together with the device current upon application of a magnetic field.

We discussed the experimental findings within the frame work of (i) the electron-hole pair model, (ii) the triplet-exciton polaron quenching model, and (iii) the bipolaron mechanism. We show that our experimental findings are in agreement only with the bipolaron mechanism.

\section{Acknowledgements}

This work was supported by NSF Grant No. ECS 04-23911.`

\bibliography{Bibliography}

\end{document}